\documentclass[aps,superscriptaddress,twocolumn,twoside,floatfix,pra,nofootinbib,a4paper]{revtex4-2}
\usepackage{times}
\usepackage{amsmath,amssymb,amsthm,mathtools,thmtools,nameref}
\usepackage{comment}
\usepackage{enumerate}
\usepackage{float}
\usepackage{xcolor}
\usepackage[colorlinks=true,linkcolor=blue,citecolor=magenta,urlcolor=blue]{hyperref}
\usepackage[framemethod=default]{mdframed}
\usepackage{csvsimple}
\usepackage[mode=tex]{standalone}
\usepackage[capitalize]{cleveref}

\usepackage{pgfplots}
\pgfplotsset{compat=1.18} 
\usepackage{tikz}
\usepackage{quantikz}

\newcommand{\beq}{\begin{equation}}
\newcommand{\eeq}{\end{equation}}

\def\>{\rangle}
\def\<{\langle}
\def\a{\alpha}
\newcommand{\bes} {\begin{subequations}}
\newcommand{\ees} {\end{subequations}}

\newcommand{\cU}{\mathcal{U}}
\newcommand{\cV}{\mathcal{V}}
\newcommand{\tcU}{\tilde{\mathcal{U}}}
\newcommand{\tcV}{\tilde{\mathcal{V}}}
\newcommand{\rtcV}{\rho_{\tcV}}
\newcommand{\rcV}{\rho_\cV}
\newcommand{\sk}{\sigma_k}
\newcommand{\tsk}{\tilde{\sk}}
\newcommand{\cF}{\mathcal{F}}
\newcommand{\commentout}[1]{}

\global\long\def\Tr{\operatorname{Tr}}%

\begin{document}
\title{Quantum Fourier Transform using Dynamic Circuits}
\author{Elisa~B\"aumer}
\email{eba@zurich.ibm.com}
\affiliation{IBM Quantum, IBM Research -- Zurich, 8803 R\"uschlikon, Switzerland}
\author{Vinay~Tripathi}
\affiliation{Department of Physics \& Astronomy, and Center for Quantum Information Science \& Technology, University of Southern California,
Los Angeles, California 90089, USA}
\author{Alireza Seif}
\affiliation{IBM Quantum, IBM T.J. Watson Research Center, Yorktown Heights, NY 10598, USA}
\author{Daniel Lidar}
\affiliation{Departments of Electrical \& Computer Engineering, Chemistry, Physics \& Astronomy, and Center for Quantum Information Science \& Technology, University of Southern California, Los Angeles, California 90089, USA}
\author{Derek~S.~Wang}
\affiliation{IBM Quantum, IBM T.J. Watson Research Center, Yorktown Heights, NY 10598, USA}

\begin{abstract}
In dynamic quantum circuits, classical information from mid-circuit measurements is fed forward during circuit execution. This emerging capability of quantum computers confers numerous advantages that can enable more efficient and powerful protocols by drastically reducing the resource requirements for certain core algorithmic primitives. In particular, in the case of the $n$-qubit quantum Fourier transform followed immediately by measurement, 
the scaling of resource requirements is reduced from $O(n^2)$ two-qubit gates in an all-to-all connectivity in the standard unitary formulation to $O(n)$ mid-circuit measurements in its dynamic counterpart without any connectivity constraints. Here, we demonstrate the advantage of dynamic quantum circuits for the quantum Fourier transform on IBM's superconducting quantum hardware with certified process fidelities of $>50\%$ on up to $16$ qubits and $>1\%$ on up to $37$ qubits, exceeding previous reports across all quantum computing platforms. These results are enabled by our contribution of an efficient method for certifying the process fidelity, as well as of a dynamical decoupling protocol for error suppression during mid-circuit measurements and feed-forward within a dynamic quantum circuit that we call ``feed-forward-compensated dynamical decoupling" (FC-DD). Our results demonstrate the advantages of leveraging dynamic circuits in optimizing the compilation of quantum algorithms.
\end{abstract}

\maketitle

Quantum computations have the potential to be more powerful than classical calculations for certain applications~\cite{NielsenChuang}. However, especially for calculations that do not utilize quantum effects, such as the addition of two bits, classical computations are likely to be faster and more reliable. Exploiting these trade-offs has resulted in the development of dynamic quantum circuits, namely quantum circuits that collect classical information from mid-circuit measurements, classically process the results, and feed-forward operations within a single shot of the circuit. A flagship application of dynamic circuits is active quantum error correction, a likely requirement of large-scale quantum computations \cite{shor1997faulttolerant,Terhal2015}.
Other applications of dynamic circuits include the generation of long-range entanglement in shallow quantum circuits used, e.g., for measurement-based quantum computations \cite{Jozsa2005,Raussendorf2003}, state preparation~\cite{buhrman2023state,malz2024prep} or the observation of phase transitions \cite{raussendorf2005longrange,piroli2021locc,kim2021measurement,tantivasadakarn2021long,kim2021measurement, tantivasadakarn2023hierarchy}, resulting in experimental studies that have demonstrated advantages with dynamic circuits for comparatively simple state preparations or gate teleportation~\cite{Corcoles2021,Fossfeig2023,Iqbal2023,moses2023race,smith2023akltstate,Cao2023,baumer2023efficient,chen2023nishi}. 

These studies typically do not explore a crucial capability of dynamic circuits: the semi-classical fan-out gate, which operates on many qubits at the same time after conditioning on the same classical output. Importantly, as the mid-circuit-measured qubits collapse to a classical state, the semi-classical fan-out gate can replace conditioned gates applied at the end of an algorithm just before the measurement of the conditioned qubit. It is then straightforward to see that the outcomes should be the same using the principle of deferred measurement, which states that any measurement can be postponed to the end of a quantum algorithm without affecting the algorithm's outcome~\cite{NielsenChuang}. Although the requirement that the measurement must immediately follow the semi-classical fan-out gate limits some applications, there do exist useful subroutines to which this idea can be applied. One such example is the quantum Fourier transform followed immediately by a measurement~\cite{Griffiths1996}, or ``QFT + measurement" (QFT+M), a subroutine used in both Shor's algorithm~\cite{shors1997} and in quantum phase estimation~\cite{kitaev1995qpe}, as well as in emerging applications~\cite{wright2024noisy,Paler2022}. 

Here, we experimentally demonstrate the QFT+M subroutine using dynamic circuits with certified process fidelities on up to 40 qubits using error suppression via dynamical decoupling (DD). The performance of the dynamic circuits-based implementation far exceeds that of its unitary implementation in this work and others on any quantum computing platform to date. To realize this comparison, we contribute an efficient method of certifying the process fidelity, as well as a DD protocol for mid-circuit measurements and feed-forward within a dynamic circuit. This demonstration of the advantage of dynamic circuits for a core quantum algorithmic primitive promises broader potential for compiling larger-scale quantum algorithms with dynamic circuits.

\begin{figure}[htb]
\includegraphics[width=1.0\columnwidth]{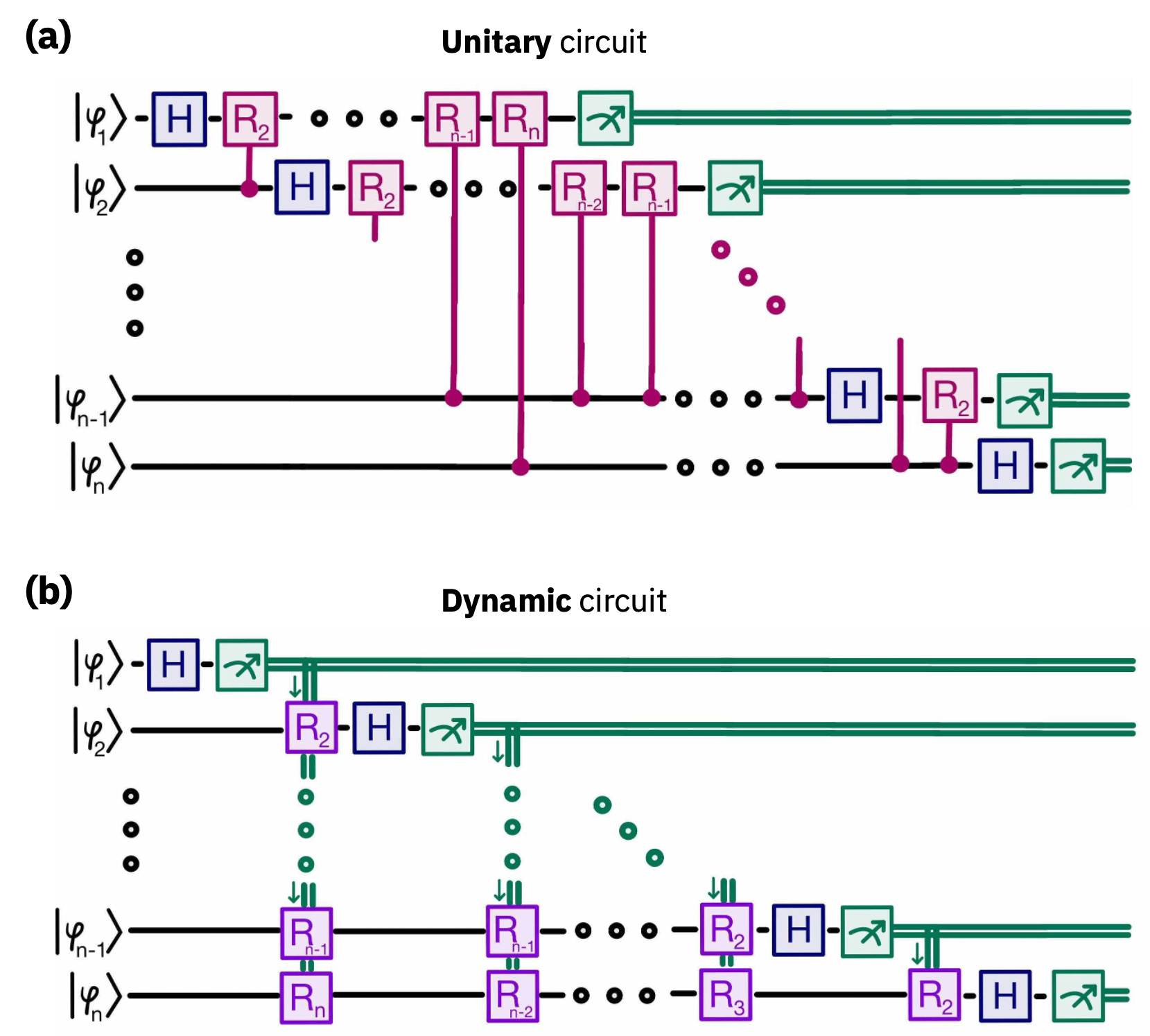}
\caption{The quantum Fourier transform with measurements implemented using 
(a) a unitary circuit and (b) a dynamic circuit, with $R_k = \begin{pmatrix} &1 &0\\ &0 &\mathrm{e}^{2\pi i /2^k}\end{pmatrix}$.
} 
\label{fig:circuit}
\end{figure}

\begin{figure*}[htb]
\centering
\includegraphics[width=1.9\columnwidth]{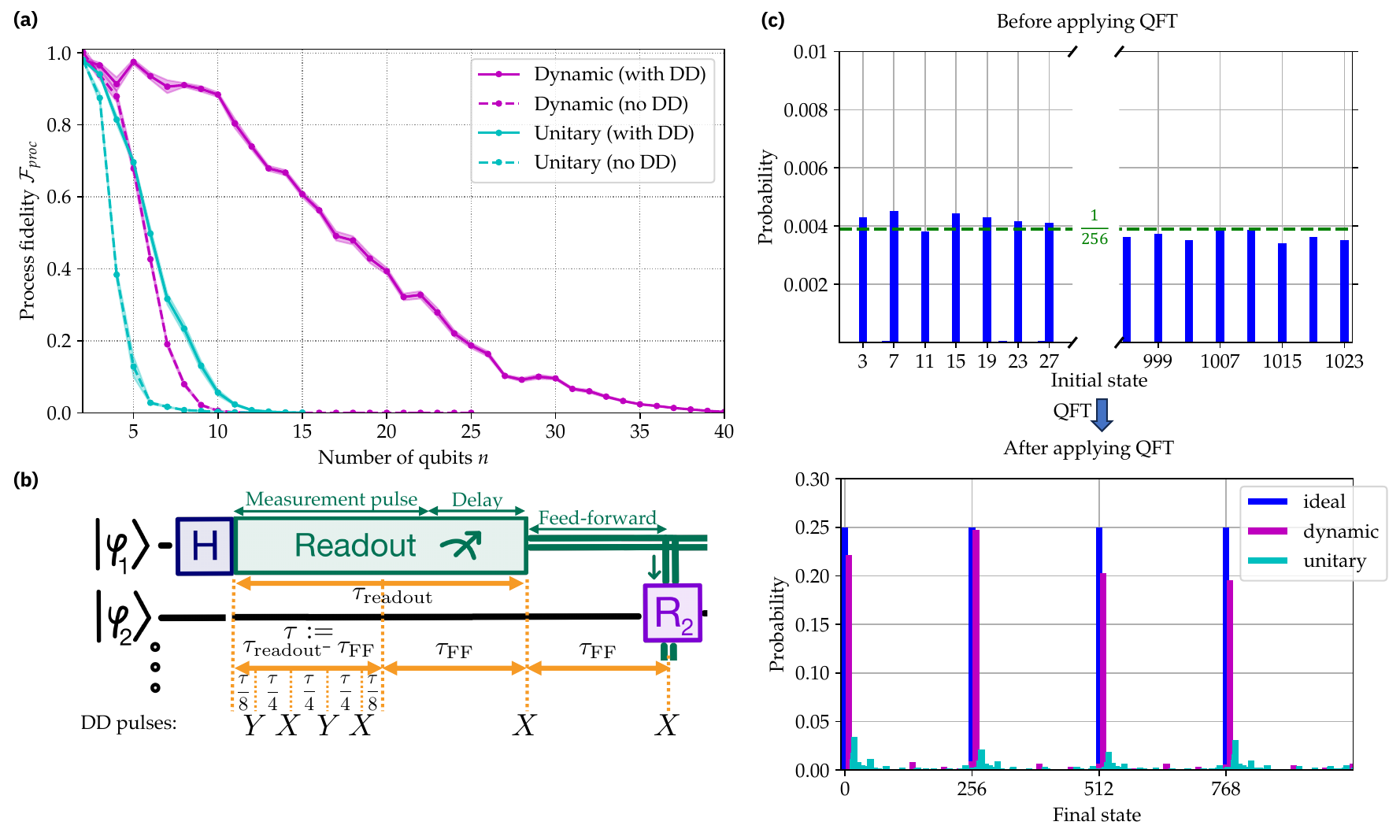}
\caption{Demonstration of the advantage of the dynamic circuit implementation of QFT+M versus the unitary version. (a) Process fidelity $\mathcal{F}_\mathrm{proc}$ of QFT+M with and without DD on \texttt{ibm\_kyiv}. All curves include readout mitigation on the final readouts via the \texttt{mthree} method~\cite{Nation2023mthree}. Shading corresponds to error bars bootstrapped over shots and input bitstrings. (b) DD scheme applied to each qubit (except the one with readout) following the Hadamard gate (H) throughout the readout with length $\tau_\mathrm{readout}=1244$~ns and feed-forward with time $\tau_\mathrm{FF}=653$~ns. No pulses can be applied during the feed-forward time, so we insert an X2 sequence with uniform spacing on either end, cutting into the readout time by $\tau_\mathrm{FF}$. In the remaining $\tau:=\tau_\mathrm{readout}-\tau_\mathrm{FF}=591$~ns idle period, we insert a uniformly spaced XY4 sequence. (c) To visualize the difference in QFT+M with the unitary vs. dynamic implementation, we first prepare a periodic state over 10 qubits, $\sum_{k=0}^{N/4-1} \ket{3+4k}$ for $N=2^{10}=1024$, where we expect the probability of each state $\ket{3+4k}$ to be $1/256$ with variations largely due to shot noise. After applying QFT, we expect 4 peaks with probability 1/4 each and observe stronger agreement for the implementation with dynamic circuits.}
\label{fig:results}
\end{figure*}

\textit{Theory}.---
As shown in \cref{fig:circuit}(a), the traditional unitary QFT+M circuit requires $O(n^2)$ two-qubit gates with all-to-all connectivity between $n$ qubits. Transpiling to device architectures with limited connectivity requires larger gate counts to mediate SWAPs \cite{moore2003generic,fowler2004implementation} that can be traded off for increased depth \cite{Park2023Reducing}, although these increased resources can then be compensated with approximate compilation~\cite{Barenco_1996,hales2000}. Note that the controlled phase gates $R_k$ are symmetric, therefore allowing the upper or lower qubits to play the role of control or target qubits interchangeably. According to the principle of deferred measurement, quantum gates commute with measurements when the qubit being measured is the control qubit. This can be applied to replace conditional quantum operations with classically controlled operations~\cite{NielsenChuang}. The result is the corresponding dynamic circuit for QFT+M as shown in \cref{fig:circuit}(b).
Crucially, this formulation does not have qubit connectivity requirements and the number of mid-circuit measurements with feed-forward scales only as $O(n)$.

To quantify the performance of QFT+M, we present an efficient method for certifying the process fidelity for a noisy unitary followed by a projective measurement. While there have been proposals for benchmarking the success probability of such a process~\cite{qedcbenchmark}, these measures are based on normalizing the classical fidelity and therefore do not define a proper quantum fidelity measure. 
In \cref{app:fidelity}, we derive a general expression for the process fidelity of a quantum channel $\cV = \Pi\circ \cU$ comprising a unitary operation $\cU$ followed by a measurement $\Pi$. This leads to the following expression for the fidelity of its noisy realization $\tilde{\cV} = \Pi\circ\tcU$, e.g., the noisy QFT+M:
\begin{align}
    \mathcal{F}_{\mathrm{proc}}(\cV,\tilde{\cV}) &= \left[\frac{1}{2^n}\sum_{k=1}^{2^n}\sqrt{ \text{Pr}\left(k| \tilde{\mathcal{QFT}}(\sigma_k^\ast)\right)}\right]^2,
\end{align}
where $\tilde{\mathcal{QFT}}$ describes the noisy QFT, $\sigma_k^\ast:=QFT^\dag\ket{k}\!\!\bra{k}QFT$ the noise-free inverse-QFT applied to the computational basis state $\ket{k}$ (binary representation of $k$), and $\text{Pr}(x|\rho)$ the probability of measuring the output $x$ given a state $\rho$. Therefore, a practical scheme to evaluate this expression is as follows: prepare the initial states $\sigma_k^\ast$, apply the noisy $\tilde{\mathcal{QFT}}$, and combine the probabilities to measure the respective outputs $k$. This resembles circuit mirroring \cite{proctor2022,hines2023scalable}, but significantly, one can prepare the initial states $\sigma_k^\ast$ with high fidelity as $QFT^\dag \ket{k} = \bigotimes_{l=1}^n R_Z(-\pi k/2^l,l) H^{\otimes n}\ket{0}$, where the $R_Z(\phi,l)$ gate with angle $\phi$ on qubit $l$ can be implemented virtually and hence noiselessly~\cite{McKay2017VirtualZ}. The scheme is efficient since, despite the fact that the number of different initial states $\sigma_k^\ast$ is $2^n$, we can sample $m$ different values $\{k_l\}_{l=1..m}$, $k_l \in \{1, .., 2^n\}$ and estimate the process fidelity as
\begin{align}
\mathcal{F}_{\mathrm{proc}}(\cV,\tilde{\cV}) &\approx \frac{m}{m-1} \left[\frac{1}{m}\sum_{l=1}^m\sqrt{ \text{Pr}\left(k_l|  \tilde{\mathcal{QFT}}(\sigma_{k_l}^\ast)\right)}\right]^2 \nonumber \\
& \quad - \frac{1}{m(m-1)}\sum_{l=1}^m \text{Pr}\left(k_l|  \tilde{\mathcal{QFT}}(\sigma_{k_l}^\ast)\right).
\end{align}
See \cref{app:fidelity} for details.

\textit{Results without DD}.---
In \cref{fig:results}(a), we compare the process fidelity $\mathcal{F}_{\mathrm{proc}}$ of QFT+M on one of IBM's superconducting quantum processors, \texttt{ibm\_kyiv}, for the unitary and dynamic circuits shown in \cref{fig:circuit}. The experimental details can be found in \cref{app:expdetails}. Comparing the ```Dynamic (no DD)'' and ```Unitary (no DD)'' curves, we see that the dynamic circuit implementation already outperforms the unitary one. However, without error suppression, the fidelities drop below $1\%$ for $n>9$ qubits, and the performance difference becomes negligible. To improve performance, we next consider the incorporation of error suppression. As the idling time scales quadratically with $n$, we expect to see a drastic increase in $\mathcal{F}_{\mathrm{proc}}$ when applying dynamical decoupling (DD).

\textit{Results with DD}.---
Dynamical decoupling~\cite{Viola1998,Khodjasteh:2005xu,Uhrig:2007qf}, whose origins are in early spin-echo experiments~\cite{CP1954,CPMG1958}, is a commonly used quantum error suppression method that has been used for years on various experimental quantum platforms~\cite{Suter:2016aa}. In essence, DD involves applying a sequence of pulses from a group of unitary operations that group-average the system-bath interaction; for every such interaction, a sequence can be found for which this average is zero~\cite{Zanardi:1999fk,Viola:99}. In particular, in the single-qubit setting, this holds for the Pauli group, and the result is the well-known XY4 sequence. 
DD has been applied to suppress decoherence \cite{Pokharel2018, Shirizly2024, souza2020process,gautamProtectionNoisyMultipartite2021,DD-survey,tong2024empirical} and crosstalk errors~\cite{Tripathi2022,Zhou2023,seif2403.06852,evert2024syncopated}, improve
quantum volume~\cite{jurcevicDemonstrationQuantumVolume2021} and algorithmic fidelity~\cite{raviVAQEMVariationalApproach2021,Pokharel:2024aa}, culminating in recent algorithmic quantum speedup demonstrations~\cite{pokharel2022demonstration,singkanipa2024demonstration}. 

A major challenge in using DD is that pulse imperfections can significantly impact performance. Thus, a careful choice of the DD sequence is necessary. 
For the unitary implementation, we use the universally robust (UR$p$) sequence family~\cite{Genov2017}, which was designed to suppress pulse axis and angle errors. It suppresses pulse errors up to $O(\epsilon^{p/2})$ using $p$ pulses, where $\epsilon$ is the nominal infidelity~\cite{Genov2017}. Varying $p$, and also comparing with other sequences from the DD survey~\cite{DD-survey}, we found that UR10 performed optimally in the setting of our experiments when inserted into the idle periods of the unitary circuit shown in \cref{fig:circuit}(a). This strategy is similar to the one used in most other recent DD experiments referenced above. However, a different strategy is needed for our dynamic circuits because they incorporate feed-forward; the corresponding DD scheme is illustrated in \cref{fig:results}(b), and we next explain its design. 

First, we note that current experimental constraints prevent the application of DD pulses immediately after the measurement pulse during the feed-forward operation with length $\tau_{\text{FF}}= 653$~ns, which is the time required to logically process mid-circuit measurements and feed-forward a quantum operation. Therefore, to maximize the suppression of noise during this feed-forward time with two uniformly spaced X pulses (the X2 sequence)~\cite{CP1954,CPMG1958}, we apply an X pulse at the beginning and the end of the feed-forward time. Because these pulses must be uniformly spaced, we do not insert any pulses for $\tau_{\text{FF}}$ into the preceding readout time $\tau_{\text{readout}} = 1244$~ns, which consists of a measurement pulse of $781$~ns followed by a delay of $463$~ns.  In the remaining time of the readout pulse $\tau :=\tau_{\text{readout}}-\tau_{\text{FF}} = 591$~ns, the uniformly spaced XY4 \cite{XY4_paper} sequence was inserted. We call this sequence feed-forward-compensated DD (FC-DD).

When comparing the ``no DD'' curves with the ``Dynamic (with FC-DD)'' and ``Unitary (with DD)'' curves, the fidelities for both the unitary and dynamic circuit implementations drastically increase with the inclusion of DD. Most notably, the fidelities observed for the dynamic circuit implementation with FC-DD are larger than those reported on any platform thus far~\cite{weinstein2001,qedcbenchmark, amico2023defining, Vorobyov2021,pfeffer2023multidimensional}. In particular, we obtain fidelities $>1\%$ for up to 37 qubits, as compared to up to 11 qubits with unitary circuits. The performance of the unitary circuits is comparable to the demonstrations of QFT in recent benchmarking works~\cite{qedcbenchmark, amico2023defining} on IBM's quantum hardware. To visualize the improvement of QFT+M with dynamic circuits, we prepare a simple periodic state $\sum_{k=0}^{2^n/4-1} |3+4k\rangle$ on $n=10$ qubits in \cref{fig:results}(c) and compare the ideal, unitary, and dynamic circuit implementation of QFT+M applied to that state. While the implementation with dynamic circuits yields a sharply peaked distribution that resembles that of the ideal case, the unitary circuits result in a flatter distribution characterized by broad peaks. This difference is expected given that the fidelities with DD are $6\%$ for unitary circuits and $88\%$ for dynamic circuits with 10 qubits; see \cref{fig:results}(a). 

A different performance metric for the dynamic circuit implementation with FC-DD---the ``average result fidelity" with plurality vote---is reported in \cref{app:resultfidelity}, and we similarly find that the present implementation outperforms the previous ones.

\textit{Conclusion and Outlook}.--- 
In this work, we demonstrate the advantage of dynamic quantum circuits, particularly for QFT+M, on IBM's superconducting quantum hardware. As first described in Ref.~\cite{Griffiths1996},  by leveraging the capability of dynamic circuits to perform mid-circuit measurements and feed-forward operations, a reduction in scaling of the required resources can be achieved from $O(n^2)$ two-qubit gates to $O(n)$ mid-circuit measurements with no connectivity constraints. Our experiments show $\mathcal{F}_\mathrm{proc}>50\%$ on up to $16$ qubits and $>1\%$ on up to $37$ qubits, setting new benchmarks across quantum computing platforms. This advancement is enabled by our efficient certification of the process fidelity and a context-aware DD protocol for dynamic circuits. 

We emphasize that the fidelities are significantly improved in the presence of a tailored DD sequence---FC-DD---that leverages the fact that the qubits are not coupled and, therefore, suffer little crosstalk noise. Although this assumption may not hold in some applications, we stress that useful cases do exist, e.g., when implementing dynamic circuits with ancilla qubits or using multiple registers such as in Shor's algorithm. Moreover, future generations of superconducting devices with tunable couplers~\cite{Stehlik2021TunableCoupler} may suffer drastically less nearest-neighbor crosstalk, suggesting even stronger potential for advantage with dynamic circuits in such devices.

Looking ahead, the successful application of dynamic circuits to the QFT+M subroutine suggests broader potential for optimizing and enabling quantum algorithms. Additionally, refining error suppression techniques and developing error mitigation methods for mid-circuit measurement and feed-forward operations in dynamic circuits, such as adapting measurement-based probabilistic error cancellation~\cite{gupta2023probabilistic}, will be crucial for unlocking their full potential.

\acknowledgements
    We thank Diego Ristè, Maika Takita, Edward H. Chen, Oles Shtanko, Luke Govia, Jay Gambetta, Emily Pritchett, and Sarah Sheldon for valuable discussions and support. The research of VT and DL was supported by the ARO MURI grant W911NF-22-S-0007. AS was sponsored by the Army Research Office  under Grant Number W911NF-21-1-0002. The views and conclusions contained in this document are those of the authors and should not be interpreted as representing the official policies, either expressed or implied, of the Army Research Office or the U.S. Government. The U.S. Government is authorized to reproduce and distribute reprints for Government purposes notwithstanding any copyright notation herein.

\appendix
\newpage

\section{Fidelity estimation}
\label{app:fidelity}
\subsection{Process fidelity of a noisy unitary followed by a measurement}

Consider a projective measurement in the $Z$-basis described by $\Pi (\rho) = \sum_k \Pi_k(\rho) = \sum_k \pi_k \rho \pi_k$ (where $\pi_k = \ket{k}\!\!\bra{k}$ and $\{\pi_k\}$ are a complete set of orthogonal projectors), an ideal unitary map $\mathcal{U}(\rho)=U\rho U^\dag$, and its noisy implementation $\tcU$. Let the ideal map be $\mathcal{V}=\Pi \circ \cU$ and its noisy version be $\tilde{\mathcal{V}} = \Pi \circ \tcU$. We can always formally decompose the noisy unitary map as $\tcU = \Lambda_U \circ\cU$ where $\Lambda_U$ is a $U$-dependent noise channel that captures all the noise acting during the noisy implementation of the ideal unitary. 

Writing $e_{ij}\equiv\ket{i}\!\!\bra{j}$, and noting that 
\bes
\begin{align}
\Pi(A) &= \sum_k \pi_k A \pi_k =  \sum_k \bra{k}A\ket{k}\pi_k \\
&= \sum_k \Tr(\pi_k A) \pi_k 
= \sum_k \< \pi_k,A\> \pi_k,
\end{align}
\ees
where $\<A,B\> = \Tr(A^\dag B) = \<B,A\>^*$ is the Hilbert-Schmidt inner product and $A,B$ are arbitrary operators, the corresponding Choi states are given by
\bes
\begin{align}
    \rtcV &\equiv (\mathbb{I} \otimes \tilde{\mathcal{V}})(\ket{\phi}\!\!\bra{\phi})=\frac{1}{d}\sum_{i,j=1}^de_{ij}\otimes \Pi[\tcU(e_{ij}) ]\\
    &=\frac{1}{d}\sum_{k=1}^d \sum_{i,j=1}^d  \Tr[\pi_k\tcU(e_{ij})] e_{ij}\otimes\pi_k\\
    &=\frac{1}{d}\sum_k\tsk\otimes\pi_k,
\end{align}
\ees
with
\bes 
\begin{align}
    \tsk &\equiv\sum_{i,j}\<\pi_k,\tcU(e_{ij})\> e_{ij}=\sum_{i,j}\<\tcU^\dag(\pi_k),e_{ij}\> e_{ij}\\
       &= \sum_{i,j}\<e_{ij},\tcU^\dag(\pi_k)\>^* e_{ij}= [\tcU^\dag(\pi_k)]^* = \tcU^T(\pi_k) ,
\end{align}
\ees
where we used $\<A,\Phi (B)\> = \<\Phi^\dag(A),B\>$ and $[\Phi(\pi_k)]^\dag = [\sum_\a K_\a \pi_k K_\a^\dag ]^\dag = \Phi(\pi_k)$ for any CP map $\Phi$ with Kraus operators $\{K_\a\}$.

Similarly, after dropping the tilde everywhere:
\beq
\label{eq:rcV}
\rcV \equiv (\mathbb{I} \otimes \mathcal{V}) (\ket{\phi}\!\!\bra{\phi}) = \frac{1}{d}\sum_k\sk\otimes\pi_k .
\eeq
Note that if $U$ is unitary then so is $U^T$ [since transposing $U^\dag U= I$ implies $U^T U^* = I$]. Thus
\beq
    \sk = \cU^T(\pi_k) 
\eeq
is a pure state. Moreover, the $\{\sk\}$ also form a (rotated) complete set of orthogonal projectors: $\sk\sigma_l= \delta_{kl}\sk$. Therefore 
\beq
\label{eq:rhoV^2}
    \rcV^2 =  \frac{1}{d^2}\sum_{k,l}\sk\sigma_l\otimes\pi_k\pi_l =  \frac{1}{d} \rcV ,
\eeq
so that $\sqrt{\rcV} = \sqrt{d}\rcV$ is the unique positive semi-definite square-root of $\rcV$.

Moreover, the results above mean that we can describe $\sk$ and $\tsk$ as the ideal and noisy transposed unitaries $\cU^T$ and $\tcU^T$, respectively, applied to the computational state $\ket{k}$. 

The process fidelity of two channels $\cV$, $\tcV$ is given by the state fidelity of their Choi states, i.e., the Uhlmann fidelity $\cF(\rcV,\rtcV) = \|\sqrt{\rcV}\sqrt{\rtcV}\|^2_1$ ($\|\cdot\|_1$ is the trace norm), which we can simplify as follows:
\bes
\begin{align}
    \cF(\rcV,\rtcV)&= \left[\Tr\sqrt{\sqrt{\rcV}\rtcV\sqrt{\rcV}}\right]^2\\
    &= \left[\Tr\sqrt{d\rcV\rtcV\rcV}\right]^2\\
    &= \frac{1}{d^2}\left[\Tr\sqrt{\sum_k\sk \tsk\sk\otimes\pi_k}\right]^2\\
    \label{eq:A13}
    &= \frac{1}{d^2}\left[\Tr\sqrt{\sum_k \Tr(\sk\tsk)\sk \otimes\pi_k}\right]^2\\
    &=  \frac{1}{d^2}\left[\Tr\sum_k \sqrt{\Tr(\sk\tsk})\sk \otimes\pi_k\right]^2
\end{align}
\ees
where in \cref{eq:A13} we used the fact that $\sk$ is pure, and in the last equality we used
\bes
\begin{align}
   & \left[\sum_k\sqrt{\Tr(\sk\tsk)}\sk\otimes \pi_k\right]^2\notag \\
   &\qquad = \sum_{k,l}\sqrt{\Tr(\sk\tsk)}\sqrt{\Tr(\sigma_l\tilde{\sigma_l})} \sk\sigma_l\otimes \pi_k\pi_{l}\\
    &\qquad = \sum_k\Tr(\sk\tsk)\sk\otimes \pi_k,
\end{align}
\ees
to determine the square root. As $\Tr(\sk \otimes \pi_k)=1$ $\forall k$, the fidelity simplifies to
\beq
\label{eq:Fproc}
    \cF_{\mathrm{proc}}(\cV,\tcV) = \left[\frac{1}{d}\sum_{k=1}^d\sqrt{\Tr(\sk\tsk)}\right]^2 .
\eeq
This can be related to the classical case, where the fidelity between two random variables $X$ and $Y$ with respective probability distributions $p=(p_{1},\ldots ,p_{n})$ and $q=(q_{1},\ldots ,q_{n})$ is given by $F(X,Y)=\left(\sum _{i}{\sqrt {p_{i}q_{i}}}\right)^{2}$. 

Now, note that 
\bes
\label{eq:tr-expand}
\begin{align}
\Tr(\sk\tsk) &= \<\sk^\dag,\tsk\> = \<\sk^\dag,\tcU^T(\pi_k)\> \\
&= \<\tcU^*(\sk^\dag),\pi_k\> = (\<\tcU^*(\sk^\dag),\pi_k\>)^*\\
& = \<\pi_k,\tcU^*(\sk^\dag)\> = \<\pi_k,\tcU^*(\sk)\> ,
\end{align}
\ees
where in the second line we used the fact that $\Tr(\sk\tsk)$ must be non-negative (hence real) since $\cF_{\mathrm{proc}}(\cV,\tcV)\ge 0$, and in the last equality we used the fact that $\sk$ is a quantum state.

Next, writing the CP map $\tcU$ in terms of its Kraus operators $\{K_\a\}$, we have 
\bes
\label{eq:tcU*}
\begin{align}
[\tcU^*(\sk)]^* &= \sum_\a (K_\a^* \sk K_\a^T)^* = \sum_\a K_\a \sk^* K_\a^\dag\\
& = \tcU(\sk^*) ,
\end{align}
\ees
where $\sk^* = \cU^\dag (\pi_k) $ is the state corresponding to the time-reversed ideal unitary $\cU$. Since $\tcU(\sk^*)$ is also a state, i.e., is positive-semi definite, it is invariant under complex conjugation. Thus, inserting $\tcU^*(\sk) = \tcU(\sk^*)$ into the fidelity and combining Eqs.~\eqref{eq:Fproc}-\eqref{eq:tcU*} finally yields
\bes
\begin{align}
    \cF_{\mathrm{proc}}(\cV,\tcV)
    &=\left[\frac{1}{d}\sum_{k=1}^d\sqrt{\bra{k} \tcU(\sk^*)\ket{k}}\right]^2\\
    &= \left[\frac{1}{d}\sum_{k=1}^d\sqrt{ \text{Pr}\left(k| \tcU(\sk^*)\right)}\right]^2,
\end{align}
\ees
where $\text{Pr}(x|\rho)$ refers to the probability of measuring the output $x$ given a state $\rho$.

This has a nice operational meaning, as it means that we can determine the fidelity of a noisy unitary that is followed by a measurement, by preparing the time-reversed initial states $\sk^* = U^\dag \pi_kU$, applying the noisy unitary $\tcU$ and combining the probabilities to measure the respective outputs $k$. This, however, requires the ability to prepare the initial states with a high fidelity.

Note that while this expression superficially resembles the one presented in Ref.~\cite{qedcbenchmark}, the sum here is over the square roots of the probabilities of the correct outputs for different input states. In contrast, in Ref.~\cite{qedcbenchmark}  the sum is over the square roots of the fidelities of different outputs for a fixed input state.

\subsection{Process fidelity of the noisy QFT+M}
Let us apply the above fidelity estimation to the case where the ideal unitary operation is the quantum Fourier transform, $\cU(\rho)=\mathcal{QFT}(\rho)=QFT \cdot \rho \cdot QFT^\dag$, and define the quantum Fourier transform followed by measurement as $\text{QFT+M}:=\cV=\Pi \circ \mathcal{QFT}$, with its noisy version $\tilde{\cV} = \Pi \circ \Lambda \circ \mathcal{QFT} = \Pi \circ \tilde{\mathcal{QFT}} $. We can prepare the initial states $\sigma_k^\ast = QFT^\dag\ket{k}\!\!\bra{k}QFT$ with high efficiency and fidelity as
\bes
\begin{align}
    QFT^\dag \ket{k} &= \frac{1}{2^n}\sum_{j=0}^{2^n-1}e^{-2\pi i k j/2^n} \ket{j} \\
    &= \bigotimes_{l=1}^n R_Z(-\pi k/2^l,l) H^{\otimes n}\ket{0},
\end{align}
\ees
which means that we simply need to apply single qubit rotations to the corresponding qubits.

As the number of different initial states $\sigma_k^\ast$ increases exponentially, we can sample $m$ different values $\{k_l\}_{l=1..m}$, $k_l \in \{1, .., 2^n\}$. 
Note that using the square of sample average (denoted by an overline) as an estimator for $E[\sqrt{X}]^2$ has an $O(1/m)$ bias.  This can be seen by noting that $E[(\bar{\sqrt{X}})^2] = E[(\frac{1}{m}\sum_i\sqrt{X_i})^2] = E[\frac{1}{m^2}(\sum_{i\neq j} \sqrt{X_iX_j}+ \sum_i X_i] = \frac{m-1}{m}E[\sqrt{X}]^2 + \frac{1}{m}E[X]$. However, we can construct an unbiased estimator of $E[\sqrt{X}]^2$ by using $\frac{m}{m-1} (\bar{\sqrt{X}})^2 - \frac{1}{m-1} \bar{X}$, and estimate process fidelity via
\bes
\begin{align}
    \mathcal{F}_{\mathrm{proc}}(\cV,\tilde{\cV}) &= \left[\frac{1}{d}\sum_{k=1}^{2^n}\sqrt{ \text{Pr}\left(k| \tilde{\mathcal{QFT}}(QFT^\dag\ket{k}\!\!\bra{k}QFT)\right)}\right]^2\\
    &\approx \frac{m}{m-1} \left[\frac{1}{m}\sum_{l=1}^m\sqrt{ \text{Pr}\left(k_l|  \tilde{\mathcal{QFT}}(\sigma_{k_l}^\ast)\right)}\right]^2 \notag \\
    &\qquad - \frac{1}{m(m-1)}\sum_{l=1}^m \text{Pr}\left(k_l|  \tilde{\mathcal{QFT}}(\sigma_{k_l}^\ast)\right).
\end{align}
\ees
Note that in practice, the correction of the second term is negligibly small.
\section{Experimental details}
\label{app:expdetails}
\begin{figure*}[htb]
\includegraphics[width=1.8\columnwidth]{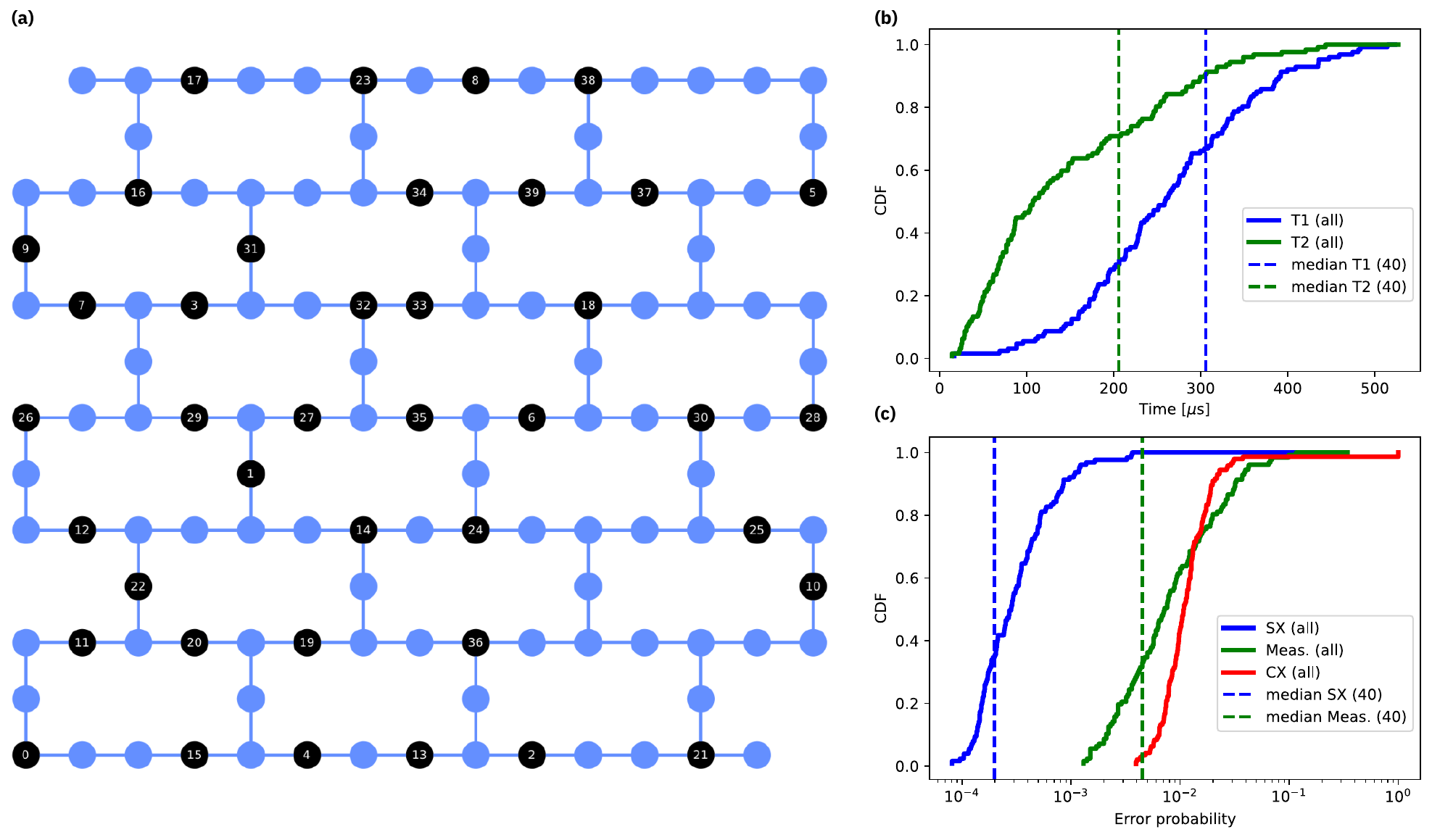}
\caption{Experimental implementation details. In (a), we show the device layout of \texttt{ibm\_kyiv}, with the 40 qubits chosen for our dynamic circuits experiments marked in black. In (b) and (c), we plot the cumulative distribution of T1 and
T2 coherence times, the single qubit gate (SX), readout (Meas.) and two qubit entangling gate (CX) error rates of all qubits on the device, as well as the corresponding median values of the chosen 40 qubits (CX gates were not used in the dynamic circuit implementation).}
\label{fig:qubitdata}
\end{figure*}

We perform all experiments on \texttt{ibm\_kyiv}, a 127-qubit superconducting quantum processor. The qubits chosen for our dynamic circuits experiments are indicated in \cref{fig:qubitdata}(a). The cumulative distribution of the device's T1 and T2 coherence times, as well as of the different error rates are shown in \cref{fig:qubitdata}(b)-(c), indicating also the corresponding median values of the 40 qubits chosen for our dynamic circuit experiments. For each $n$, we sampled $m=20$ random bit strings $k$, corresponding to the initial states $\sigma_k^\ast$, and measured 2000 shots each.

\section{Average result success probability}
\label{app:resultfidelity}
\begin{figure*}[htb]
\includegraphics[width=1.2\columnwidth]{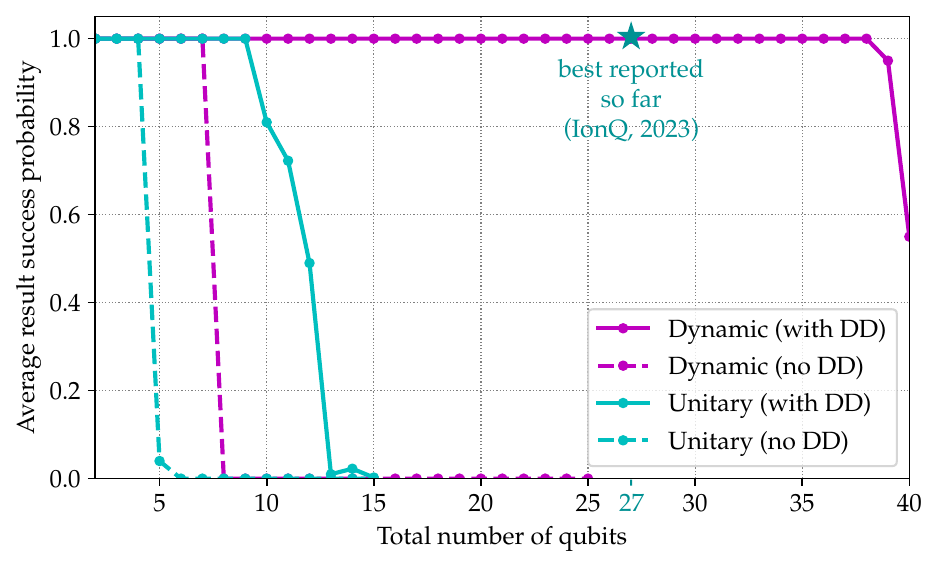}
\caption{Demonstration of the advantage of the dynamic circuit implementation of QFT+M versus the unitary version for another metric: the average result success probability.}
\label{fig:pluralityvote}
\end{figure*}

Ref.~\cite{chen2023benchmarking} reported a so-called ``average result fidelity" of close to $100\%$ for the QFT on up to 27 qubits. These results rely on plurality voting as a post-processing technique, which assumes that the ideal result is a single bitstring. For this reason, the ``average result fidelity" is not a valid \textit{fidelity} measure and we describe it instead as the ``average result success probability." Nevertheless, for the sake of comparison, we applied a plurality vote to our results to evaluate this quantity as well. The results are shown in \cref{fig:pluralityvote}, where we report an average success probability of $100\%$ for up to 38 qubits. The deterioration of our unitary results compared to the previous best reported results using the above metric is due to an all-to-all connectivity which reduces the requirement for numerous SWAP operations, a limitation that is particularly pronounced in our superconducting systems and negatively impacts the performance.

\end{document}